\documentclass[twocolumn, twocolappendix]{aastex631}

\received{November 16, 2022}
\revised{December 19, 2022}
\accepted{December 20, 2022}

\submitjournal{ApJ}

\shorttitle{The \textit{Hubble Space Telescope} UV Legacy Survey of Galactic Globular Clusters. XXIV.}

\shortauthors{Libralato et al.}

\usepackage{xspace}
\usepackage{amsmath}
\usepackage{rotating}
\usepackage{threeparttable}
\usepackage{enumerate}
\usepackage{soul}

\newcommand{\hstfull}{\textit{Hubble Space Telescope}\xspace}
\newcommand{\hst}{\textit{HST}\xspace}

\begin{document}

\title{The \textit{Hubble Space Telescope} UV Legacy Survey of  Galactic Globular Clusters. XXIV. Differences in internal kinematics of multiple stellar populations}

\correspondingauthor{Mattia Libralato}
\email{libra@stsci.edu}

\author[0000-0001-9673-7397]{Mattia Libralato}
\affil{AURA for the European Space Agency (ESA), Space Telescope Science Institute, 3700 San Martin Drive, Baltimore, MD 21218, USA}

\author[0000-0003-2742-6872]{Enrico Vesperini}
\affil{Department of Astronomy, Indiana University, Bloomington, IN 47405, USA}

\author[0000-0003-3858-637X]{Andrea Bellini}
\affil{Space Telescope Science Institute 3700 San Martin Drive, Baltimore, MD 21218, USA}

\author[0000-0001-7506-930X]{Antonino P. Milone}
\affil{Dipartimento di Fisica e Astronomia, Universit\`a di Padova, Vicolo dell'Osservatorio 3, Padova, I-35122, Italy}
\affil{INAF - Osservatorio Astronomico di Padova, Vicolo dell'Osservatorio 5, Padova, I-35122, Italy}

\author[0000-0001-7827-7825]{Roeland P. van der Marel}
\affil{Space Telescope Science Institute 3700 San Martin Drive, Baltimore, MD 21218, USA}
\affil{Center for Astrophysical Sciences, The William H. Miller III Department of Physics \& Astronomy, Johns Hopkins University, Baltimore, MD 21218, USA}

\author[0000-0002-9937-6387]{Giampaolo Piotto}
\affil{Dipartimento di Fisica e Astronomia, Universit\`a di Padova, Vicolo dell'Osservatorio 3, Padova, I-35122, Italy}
\affil{INAF - Osservatorio Astronomico di Padova, Vicolo dell'Osservatorio 5, Padova, I-35122, Italy}

\author[0000-0003-2861-3995]{Jay Anderson}
\affil{Space Telescope Science Institute 3700 San Martin Drive, Baltimore, MD 21218, USA}

\author[0000-0002-6054-0004]{Antonio Aparicio}
\affil{Instituto de Astrof\'isica de Canarias, Calle V\'ia L\'actea, La Laguna, Tenerife, Canary Islands, E-38205, Spain}
\affil{Departamento de Astrof\'isica, Universidad de La Laguna, Av. Astrof\'sico Francisco S\'anchez, La Laguna, Tenerife, Canary Islands, E-38206, Spain}
\affil{Department of Aerospace Engineering, Khalifa University of Science and Technology, P.O. Box 127788, Abu Dhabi, United Arab Emirates}

\author[0000-0001-9264-4417]{Beatriz Barbuy}
\affil{Universidade de S\~ao Paulo, IAG, Rua do Mat\~ao 1226, Cidade Universit\'asria, S\~ao Paulo 05508-900, Brazil}

\author[0000-0003-4080-6466]{Luigi R. Bedin}
\affil{INAF - Osservatorio Astronomico di Padova, Vicolo dell'Osservatorio 5, Padova, I-35122, Italy}

\author[0000-0002-1793-9968]{Thomas M. Brown}
\affil{Space Telescope Science Institute 3700 San Martin Drive, Baltimore, MD 21218, USA}

\author[0000-0001-5870-3735]{Santi Cassisi}
\affil{INAF - Osservatorio Astronomico di Abruzzo, Via M. Maggini, s/n, Teramo, I-64100, Italy}
\affil{INFN - Sezione di Pisa, Largo Pontecorvo 3, Pisa, I-56127, Italy}

\author[0000-0003-1149-3659]{Domenico Nardiello}
\affil{INAF - Osservatorio Astronomico di Padova, Vicolo dell'Osservatorio 5, Padova, I-35122, Italy}

\author[0000-0001-6708-4374]{Ata Sarajedini}
\affil{Department of Physics, Florida Atlantic University, Boca Raton, FL 33431, USA}

\author[0000-0001-8834-3734]{Michele Scalco}
\affil{Dipartimento di Fisica e Scienze della Terra, Università di Ferrara, Via Giuseppe Saragat 1, Ferrara, I-44122, Italy}
\affil{INAF - Osservatorio Astronomico di Padova, Vicolo dell'Osservatorio 5, Padova, I-35122, Italy}
\affil{Department of Astronomy, Indiana University, Bloomington, IN 47405, USA}

\begin{abstract}
Our understanding of the kinematic properties of multiple stellar populations (mPOPs) in Galactic globular clusters (GCs) is still limited compared to what we know about their chemical and photometric characteristics. Such limitation arises from the lack of a comprehensive observational investigation of this topic. Here we present the first homogeneous kinematic analysis of mPOPs in 56 GCs based on high-precision proper motions computed with \hstfull data. We focused on red-giant-branch stars, for which the mPOP tagging is clearer, and measured the velocity dispersion of stars belonging to first (1G) and second generations (2G). We find that 1G stars are generally kinematically isotropic even at the half-light radius, whereas 2G stars are isotropic at the center and become radially anisotropic before the half-light radius. The radial anisotropy is induced by a lower tangential velocity dispersion of 2G stars with respect to the 1G population, while the radial component of the motion is comparable. We also show possible evidence that the kinematic properties of mPOPs are affected by the Galactic tidal field, corroborating previous observational and theoretical results suggesting a relation between the strength of the external tidal field and some properties of mPOPs. Although limited to the GCs' central regions, our analysis leads to new insights into the mPOP phenomenon, and provides the motivation for future observational studies of the internal kinematics of mPOPs.\looseness=-4
\end{abstract}

\defcitealias{2022ApJ...934..150L}{L22}


\section{Introduction}\label{sec:intro}

\begin{figure*}
    \centering
    \includegraphics[width=\columnwidth]{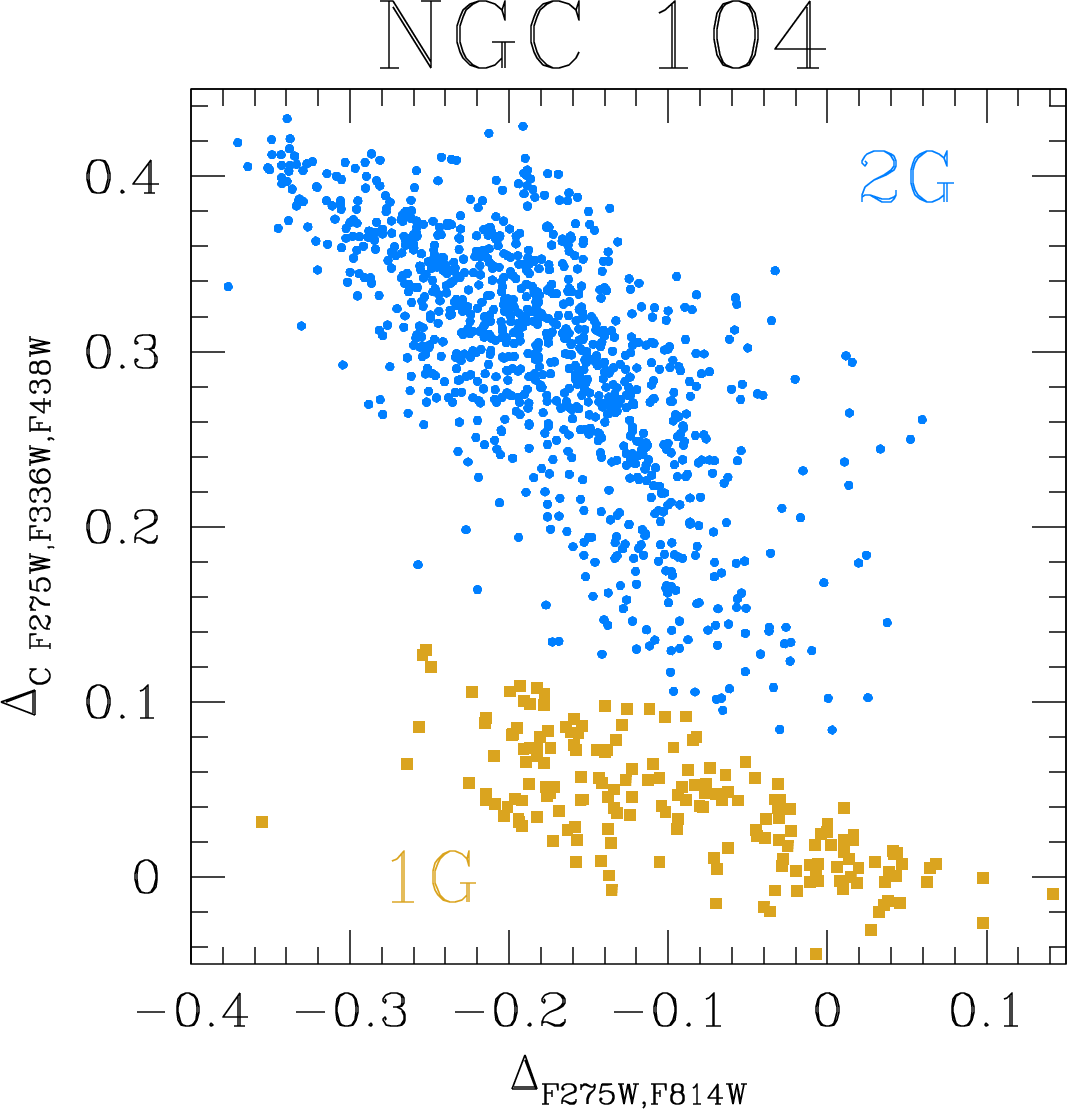}
    \includegraphics[width=\columnwidth]{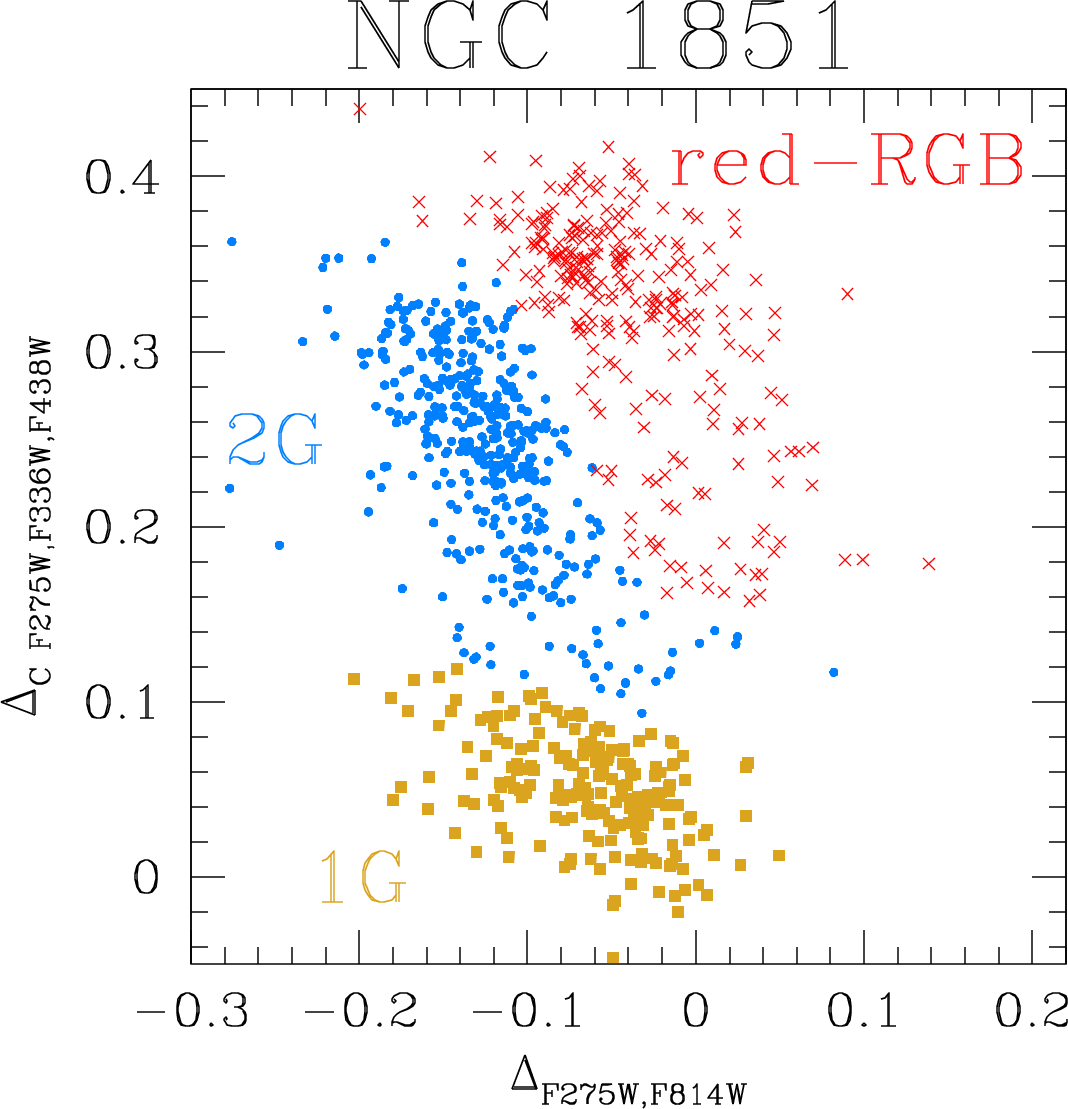}
    \caption{Examples of ``chromosome'' maps and mPOP tagging for a type-I (NGC~104; left panel) and a type-II (NGC~1851; right panel) GCs. In each plot, gold squares and blue dots represent 1G and 2G stars on the RGB, respectively. Red crosses in the right panel highlight the red-RGB stars in NGC~1851.\looseness=-4}
    \label{fig:chm}
\end{figure*}

The puzzle of the origin of the multiple stellar populations (mPOPs) in Galactic globular clusters (GCs) has been controversial since their discovery. The large amount of spectroscopic and photometric data collected so far has provided almost all observational information we know about mPOPs in GCs, but no definitive consensus has been reached yet about the formation and evolution of mPOPs \citep{2015MNRAS.454.4197R, 2018ARA&A..56...83B, 2012A&ARv..20...50G, 2019A&ARv..27....8G,2020A&ARv..28....5C,2022Univ....8..359M}. The interplay between theoretical and observational efforts has pushed the community to find new ways to constrain the origin of mPOPs in GCs. For example, this research field is progressively seeking answers by looking at young and massive clusters in other galaxies \citep[e.g.,][]{2014ApJ...797...15L,2016ApJ...829...77D,2017MNRAS.465.4159N,2019ApJ...871..140L,2019MNRAS.487.5324M,2019MNRAS.485.3076N,2020MNRAS.491..515M}. However, there is still an almost uncharted wealth of information in Galactic GCs that can enrich the observational picture of mPOPs: their internal kinematics.\looseness=-4

Here, we investigate the kinematic properties of first (1G) and second (2G) generation stars hosted in GCs. This effort focuses on red-giant branch (RGB) stars, for which the separation between different populations is clearer. We make use of the homogeneous collection of proper motions (PMs) obtained with \hstfull (\hst) data recently published by \citet[hereafter L22]{2022ApJ...934..150L} for 56 globular and one open clusters, and compare the properties of the velocity distributions of 1G and 2G stars.\looseness=-4

The outline of the paper is the following. Section~\ref{sec:data} describes the data sets used for this study, the procedure to identify the mPOPs, and how we calculated their kinematic properties. Section~\ref{sec:kin} reports our results concerning the kinematics of mPOPs; while in Section~\ref{sec:rperi} we investigate the possible dependence of the velocity anisotropy on the Galactic tidal field.\looseness=-4

\section{Data sets, multiple-population tagging and kinematics}\label{sec:data}

We made use of the PM catalogs\footnote{Catalogs are available at MAST as a High Level Science Product via \protect\dataset[10.17909/jpfd-2m08]{\doi{10.17909/jpfd-2m08}}. See also: \href{https://archive.stsci.edu/hlsp/hacks}{https://archive.stsci.edu/hlsp/hacks}.} of \citetalias{2022ApJ...934..150L}, to which we refer for a detailed description of how the PMs were computed\footnote{Since our focus is on GCs, we excluded the open cluster NGC~6791 from the investigation.}. In brief, we designed a multi-step reduction to exploit crowded regions, like the cores of GCs, and measured position and flux of sources in \hst exposures via effective point-spread function fitting. The geometric-distortion-corrected positions of each object as a function of time were then fit with a least-squares straight line, the slope of which is an estimate of the PM of the star. We cross-identified stars in our astro-photometric catalogs with those in the (pseudo) two-color diagrams known as ``chromosome maps'' of \citet{2017MNRAS.464.3636M} made for all clusters analyzed in the Treasury GO-13297 program \citep{2015AJ....149...91P}. We then applied the astro-photometric quality selections described in both papers to obtain samples of well-measured RGB stars for the mPOP tagging and their kinematic analysis.\looseness=-4

\begin{table*}[th!]
    \centering
    \caption{Median anisotropy of 1G, 2G and red-RGB stars for stars with $r > 0.6\,r_{\rm h}$ and statistical significance of the difference between the mPOP anisotropies for the various cases discussed in text. The values between brackets in the second, third and fourth columns are the number of stars used in each case. No number is provided when no stars are available, or if the anisotropy for the specific case was not computed (see text for details).}
    \label{tab:anisotropy}
    \begin{tabular}{c|c|c|c|c|c|c}
        \hline
        \hline
        & 1G ($N$) & 2G ($N$) & red-RGB ($N$) & 1G vs. 2G & 1G vs. red-RGB & 2G vs. red-RGB \\
        \hline
        & \multicolumn{3}{c|}{Median anisotropy} & \multicolumn{3}{c}{Comparison} \\
        \hline
        Entire sample & 1.02 $\pm$ 0.02 (5962) & 0.91 $\pm$ 0.01 (13884) & 0.90 $\pm$ 0.03 (1317) & 4.9$\sigma$ & 3.3$\sigma$ & 0.3$\sigma$ \\
        \hline
        age/$t_{\rm h}$$\ge$10 & 1.02 $\pm$ 0.03 (1138) & 0.95 $\pm$ 0.02 (2289) & / & 1.9$\sigma$ & / & / \\
        7$\le$age/$t_{\rm h}$$<$10 & 1.02 $\pm$ 0.06 (885) & 0.97 $\pm$ 0.02 (1745) & / & 0.8$\sigma$ & / & / \\
        age/$t_{\rm h}$$<$7 & 0.98 $\pm$ 0.03 (3939) & 0.92 $\pm$ 0.02 (9850) & 0.90 $\pm$ 0.07 (940) & 1.6$\sigma$ & 1.0$\sigma$ & 0.3$\sigma$ \\
        \hline
        $\rm R_{\rm peri} \le 3.5$ kpc & 1.01 $\pm$ 0.03 (2479) & 0.94 $\pm$ 0.02 (6929) & / & 1.9$\sigma$ & / & / \\
        $\rm R_{\rm peri} > 3.5$ kpc & 0.94 $\pm$ 0.07 (1460) & 0.91 $\pm$ 0.03 (2921) & / & 0.4$\sigma$ & / & / \\
        \hline
    \end{tabular}
\end{table*}

\begin{table*}[th!]
    \centering
    \caption{Slopes \textit{m} of the least-squares straight-line fits (in linear units of $r/r_{\rm h}$) to the anisotropy profiles of 1G, 2G and red-RGB stars, and statistical significance of the difference between the mPOP slopes for the various cases discussed in text.}
    \label{tab:slope}
    \begin{tabular}{c|c|c|c|c|c|c}
        \hline
        \hline
        & 1G & 2G & red-RGB & 1G vs. 2G & 1G vs. red-RGB & 2G vs. red-RGB \\
        \hline
        & \multicolumn{3}{c|}{Slope} & \multicolumn{3}{c}{Comparison} \\
        \hline
        Entire sample & -0.02 $\pm$ 0.02 & -0.06 $\pm$ 0.01 & -0.07 $\pm$ 0.02 & 1.8$\sigma$ & 1.8$\sigma$ & 0.4$\sigma$ \\
        \hline
        age/$t_{\rm h}$$\ge$10 & 0.02 $\pm$ 0.03 & -0.05 $\pm$ 0.01 & / & 2.2$\sigma$ & / & / \\
        7$\le$age/$t_{\rm h}$$<$10 & -0.01 $\pm$ 0.04 & -0.03 $\pm$ 0.02 & / & 0.4$\sigma$ & / & / \\
        age/$t_{\rm h}$$<$7 & -0.02 $\pm$ 0.02 & -0.06 $\pm$ 0.01 & -0.10 $\pm$ 0.03 & 1.8$\sigma$ & 2.2$\sigma$ & 1.3$\sigma$ \\
        \hline
        $\rm R_{\rm peri} \le 3.5$ kpc & -0.01 $\pm$ 0.02 & -0.06 $\pm$ 0.02 & / & 1.8$\sigma$ & / & / \\
        $\rm R_{\rm peri} > 3.5$ kpc & -0.05 $\pm$ 0.03 & -0.08 $\pm$ 0.02 & / & 0.7
$\sigma$ & / & / \\
        \hline
    \end{tabular}
\end{table*}

The reason of our choice to focus on the RGB stars is twofold. First, the mPOPs along the RGB can be identified more easily because the UV data (which provide the key filters needed for disentangling the mPOPs depending on their CNO contents) used by \citet{2017MNRAS.464.3636M} were designed to have the highest signal-to-noise ratio (and hence photometric quality) for the RGB stars \citep[see discussion in][]{2015AJ....149...91P}. Secondly, this choice allows us to compare the kinematics of stars with similar masses.\looseness=-4

\begin{figure*}[t!]
\centering
\includegraphics[width=\textwidth]{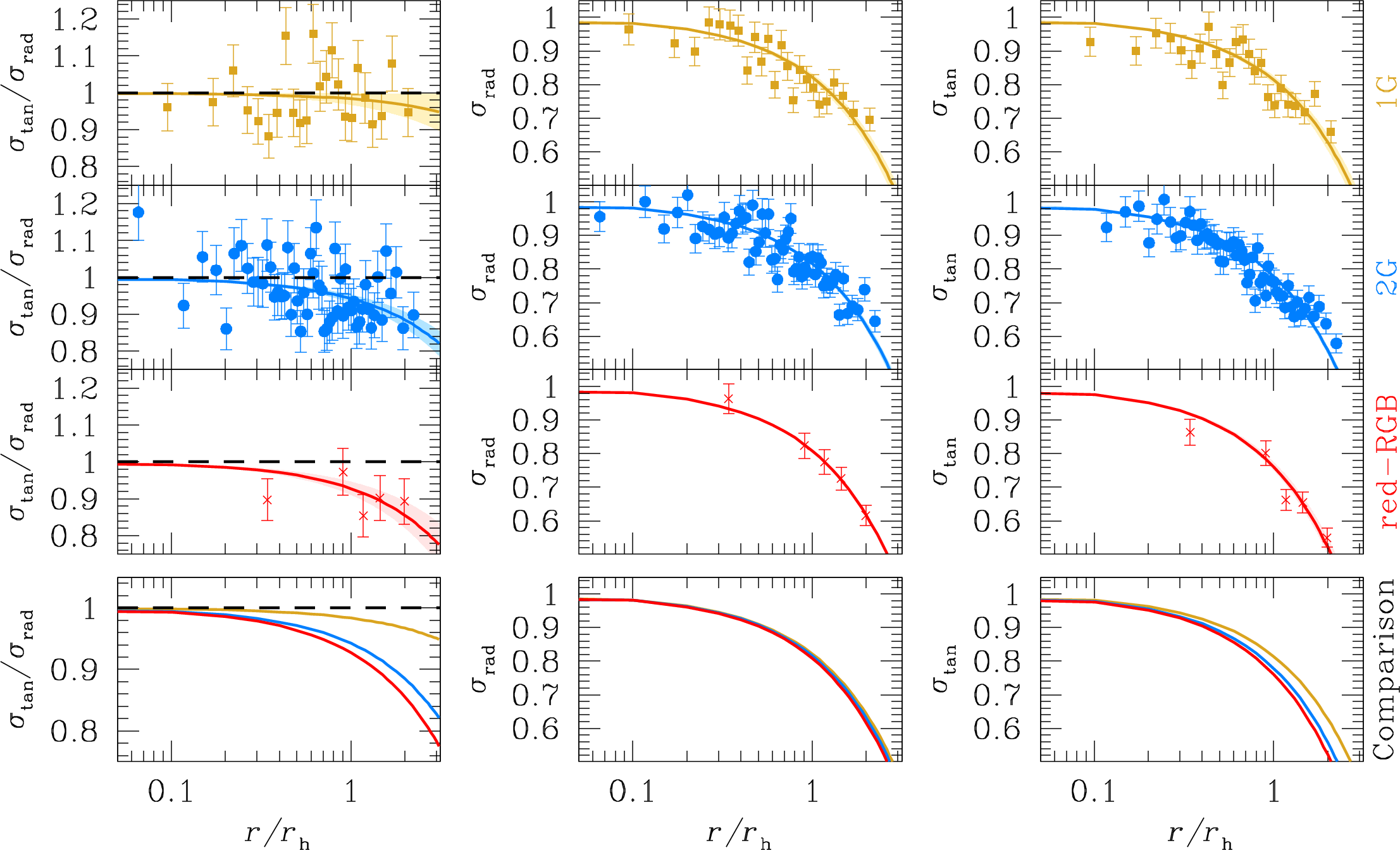}
\caption{Anisotropy (left), normalized $\sigma_{\rm rad}$ (center) and $\sigma_{\rm tan}$ (right) as a function of distance from the center of the cluster in units of $r_{\rm h}$. Gold squares (top panel), blue dots (second panel from the top) and red crosses (third panel from the top) represent the 1G, 2G and red-RGB populations, respectively. The black, dashed horizontal lines in the left panels mark the isotropic case. The solid lines in each panel, color-coded as the corresponding points, are least-squares straight-line fits  (in linear units of $r/r_{\rm h}$) to the points forced to have the ordinate equal to 1 at the center ($r/r_{\rm h} = 0$). The light-color shaded regions correspond to the 1$\sigma$ errors of the fits. The comparisons between the trends in each case are shown in the bottom panels (1$\sigma$ errors of the fits are not plotted for clarity).\looseness=-4}
\label{fig:kin1}
\end{figure*}

However, focusing on RGB stars necessarily implies small number statistics, and this poses a problem if we choose to work with each cluster separately, computing the velocity dispersions for stars in each mPOP first, and then collecting all measurements from all GCs to study the average kinematic trends of mPOPs. For clusters with different population sizes, the velocity dispersion representing the kinematics of stars at a given distance from the center of the cluster could be obtained by considering stars over a different radial interval. This could potentially introduce a bias that can wash out some of the features we are looking for. For this reason, we instead normalized positions and PMs of the RGB stars in each GC catalog by the GC's half-light radius\footnote{Cluster parameters (half-light radius, distance, half-mass relaxation time, average perigalactic distance) are taken from the GC \protect\href{https://people.smp.uq.edu.au/HolgerBaumgardt/globular/}{database} of Holger Baumgardt \citep{2017MNRAS.471.3668S,2020PASA...37...46B,2021MNRAS.505.5978V,2021MNRAS.505.5957B}. Ages are from \citet{2010ApJ...708..698D}, \citet{2014ApJ...785...21M} and \citet{2014A&A...565A..23K}. See \citetalias{2022ApJ...934..150L} for details.} ($r_{\rm h}$) and central velocity dispersion $\sigma_\mu$ (from \citetalias{2022ApJ...934..150L}), respectively. The errors on the central $\sigma_\mu$ were included in the normalized-PM error budget. This normalization allowed us to jointly compare pairs of stellar positions and PMs from all clusters at once, thus increasing the number of data points that can be used to study the kinematics of each mPOP without the drawback discussed before.\looseness=-4

\citet{2017MNRAS.464.3636M} classifies GCs in two main families. Most GCs belong to the type-I family, and are characterized by chromosome maps with two distinct groups\footnote{Note that 1G stars in type-I GCs present a color spread in the chromosome maps likely due to a spread in Fe of $\gtrsim$0.1 dex \citep{2019MNRAS.487.3815M,2022A&A...662A.117L,2022MNRAS.513..735L}.} made by 1G and 2G stars (left panel of Fig.~\ref{fig:chm}).\looseness=-4

The remaining clusters are instead labelled as type-II GCs. These systems present: more complex chromosome maps, where the 1G and 2G stars seem to be divided in subgroups (right panel of Fig.~\ref{fig:chm}); and split sub-giant branches and RGBs (hereafter red-RGB) clearly visible in color-magnitude diagrams (CMDs) with specific color combinations.\looseness=-4

Stars belonging to the red-RGB population are typically enriched in the overall CNO content, iron and s-elements abundances, and have their own 1G and 2G subdivision (hereafter, 1Gr and 2Gr, respectively). We refer to \citet{2019MNRAS.487.3815M} for a comprehensive description of the spectro-photometric properties of these stars. The origin of these red-RGB stars is not clear. For example, \citet{2019MNRAS.487.3815M} suggested two possible options: (i) after the gas from which the ``classical'' 1G and 2G stars formed was almost exhausted, the clusters reaccreted pristine gas that was enriched in iron by supernovae; or (ii) the type-II GCs formed within a dwarf galaxy, with 1G/2G and 1Gr/2Gr born at different times and/or places.\looseness=-4

The 1Gr/2Gr distinction is not always as clear as that between the 1G/2G stars in our chromosome maps, in particular for NGC~1851 and NGC~6715. Nevertheless, we arbitrarily divided the red-RGB group is our type-II clusters in 1Gr and 2Gr stars similarly to what done for 1G and 2G stars in Fig.~\ref{fig:chm}. The majority of the red-RGB stars belongs to the 2Gr group, and only 219 stars are part of the 1Gr group. Because the photometric tagging of 1Gr and 2Gr stars is not straightforward in our sample of type-II GCs, and given the very few 1Gr stars, we choose to analyzed the red-RGB stars as a whole\footnote{The kinematic analysis shown in Fig.~\ref{fig:kin1} provides consistent results for 1Gr, 2Gr and 1Gr+2Gr stars.}.\looseness=-4

Following this classification, we divided our samples in either two (1G and 2G for type-I GCs) or three (1G, 2G and red-RGB for type-II GCs) sub-populations. The mPOP tagging was directly performed on the chromosome map.\looseness=-4

\section{Global kinematic properties of multiple populations}\label{sec:kin}

As shown in a number of theoretical studies, the velocity anisotropy may provide various fundamental insights into the formation and evolution of GCs \citep[e.g.,][]{2016MNRAS.461..402T,2017MNRAS.471.2778B,2021MNRAS.502.4762B,2021MNRAS.504L..12P,2022MNRAS.509.3815P} and their mPOPs \citep{2019MNRAS.487.5535T,2021MNRAS.502.4290V}.\looseness=-4

We computed the normalized radial and tangential velocity dispersions ($\sigma_{\rm rad}$ and $\sigma_{\rm tan}$, respectively) in equally-populated radial bins of at least 200 stars each\footnote{For the analysis of the kinematics of red-RGB stars in GCs with different dynamical age, we made at least one radial bin with all stars at disposal when not enough stars were available.} as in Sect.~4 of \citetalias{2022ApJ...934..150L}, using a maximum-likelihood approach. The left panels of Fig.~\ref{fig:kin1} present the anisotropy as a function of radial distance from the center of the cluster in units of $r_{\rm h}$ for 1G, 2G and red-RGB stars. The solid lines in each panel, color-coded as the corresponding points, are least-squares straight-line fits to the points forced to have the ordinate equal to 1 at the center ($r/r_{\rm h} = 0$), i.e., $(\sigma_{\rm tan}/\sigma_{\rm rad})(r) = 1+m \times r/r_{\rm h}$. These fits are linear in $r/r_{\rm h}$, thus they appear curved in our plots with a logarithmic scale on the $x$ axis. The 1G stars (gold points) in our fields are isotropic even outside 1 $r_{\rm h}$, with only a marginal ($\sim$1$\sigma$) signature of a radial anisotropy in the outermost part of the field. The 2G (blue) and red-RGB (red) populations are isotropic in the center and become progressively radially anisotropic further from the GC's center. Table~\ref{tab:anisotropy} collects the median values of the anisotropy for each population for $r > 0.6\,r_{\rm h}$. This threshold was chosen as a compromise between having enough points to compute a robust median anisotropy value for each mPOP and being sufficiently far from the center of the cluster to capture indications of anisotropy. Table~\ref{tab:slope} collects the slopes of the straight-line fits. There is a clear difference in the median anisotropy between 1G and 2G stars at the $\sim$5$\sigma$ level, and between 1G and red-RGB stars at the $\sim$3$\sigma$ level. The slopes of the straight-line fits provide similar results, although the statistical significance is smaller ($\sim$2$\sigma$).\looseness=-4

These findings are in general agreement with the predictions of numerical models of the evolution of multiple-population clusters \citep[see, e.g.,][]{2021MNRAS.502.4290V}. Specifically, theoretical models explain these behaviors as the consequence of the initial spatial differences between 1G and 2G stars. Simulations usually start with a 2G population more centrally concentrated in the inner regions of a more diffuse 1G system as suggested by models of formation of mPOPs \citep{2008MNRAS.391..825D,2019MNRAS.489.3269C}. For systems starting with an isotropic velocity distribution, the anisotropy of the 2G stars is a consequence of the outward diffusion of 2G stars \citep{2016MNRAS.455.3693T,2021MNRAS.502.4290V}. For systems starting with an anisotropic velocity distribution, this difference is the result of a more rapid evolution towards a isotropic velocity distribution of 1G stars. In such case, it is also possible to find both populations to be characterized by anisotropic velocity distributions. We point out that although the difference between the anisotropy of 1G and 2G stars is small, its extent is generally consistent with that found in numerical models at the distances from the clusters' centers probed by our data.\looseness=-4

The middle and right panels show the radial profiles of the normalized $\sigma_{\rm rad}$ and $\sigma_{\rm tan}$, respectively. All the populations have similar radial velocity dispersions, while tangential velocity dispersions are larger for 1G stars than for 2G and red-RGB sources. These findings are in agreement with the theoretical predictions for which the different degrees of radial anisotropy between 1G and 2G stars is caused by a difference in the tangential component of their motions, rather than in the radial component \citep{2015ApJ...810L..13B,2021MNRAS.502.4290V}.\looseness=-4

\begin{figure*}[th!]
\centering
\includegraphics[width=\textwidth]{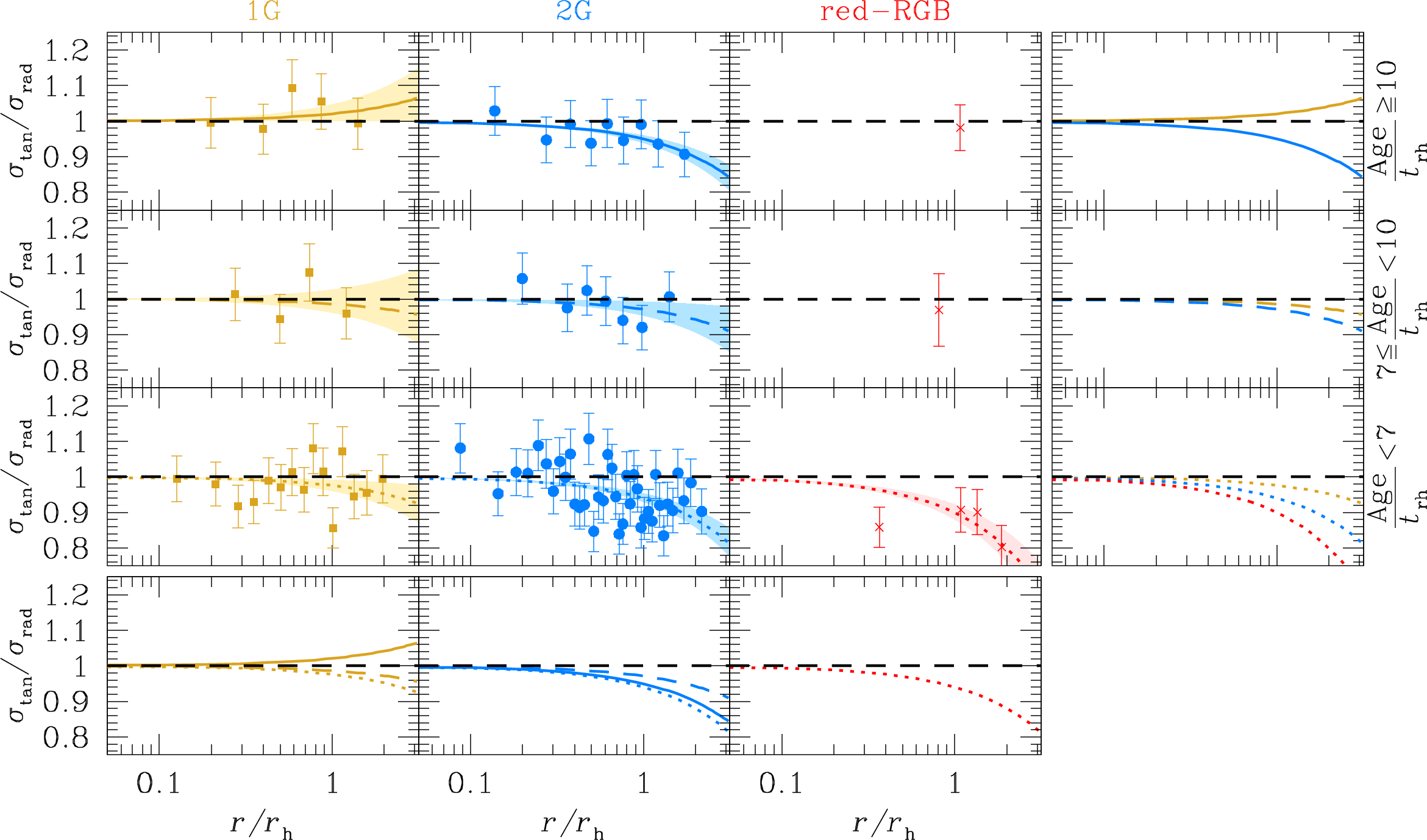}
\caption{As in the left panels of Fig.~\ref{fig:kin1}, but dividing the sample of GCs according to their dynamical age (age/$t_{\rm h}$ ratio). The first three columns show the anisotropy for 1G (gold squares), 2G (blue dots) and red-RGB (red crosses), respectively. The black, horizontal lines mark the isotropic case. The lines, colored as the point in the same plot, are a straight-line fit to the data (see details in Fig~\ref{fig:kin1}; no line was fit for the red-RGB samples with only one data point). The first three rows present the result for old, intermediate and young GCs, from top to bottom, respectively. The rightmost panels collect the straight-line fits for each population for GCs with the same the dynamical age, while the panels at the bottom show the comparison between the trends of each mPOP in clusters with different dynamical ages.\looseness=-4}
\label{fig:kin2}
\end{figure*}

In Fig.~\ref{fig:kin2}, we further explore the anisotropy of mPOPs for GCs with different dynamical ages as measured by the ratio of the GCs' ages to their half-mass relaxation time ($t_{\rm h}$). \citetalias{2022ApJ...934..150L} found that dynamically-old (age/$t_{\rm h}$$>$10) and young (age/$t_{\rm h}$$<$7) GCs are characterized by different velocity distributions at $r_{\rm h}$. We followed this same classification to better highlight differences and analogies between 1G and 2G stars. Figure~\ref{fig:kin2} presents the anisotropy as a function of distance from the center of the cluster in units of $r_{\rm h}$ for dynamically old (top), intermediate (second from the top) and young (third from the top) GCs. The bottom panels show the comparison between the trends of each mPOP in GCs with different dynamical ages, while the rightmost panels collect the straight-line fits for each population for GCs with the same the dynamical age. The trends shown in these panels are consistent with the global trends shown in Fig.~\ref{fig:kin1}, but the differences between the mPOP anisotropies are less evident (see Tables~\ref{tab:anisotropy} and \ref{tab:slope}). Most of the 1G fits are consistent with an isotropic distribution, while most of the 2G median anisotropies and slopes indicate a statistically-significant anisotropy. Our analysis suggests that, for a given mPOP, there might be kinematic differences depending on the dynamical age of the hosting cluster, but the large error bars do not allow us to draw any definitive conclusion. Larger differences might be present further from the center of the cluster, where the relaxation time is longer and fingerprints of the initial kinematic properties might still be detectable. Finally, no conclusion can be inferred for the red-RGB stars in intermediate and old GCs because of the low statistics.\looseness=-4

\begin{figure*}[ht!]
\centering
\includegraphics[width=\textwidth]{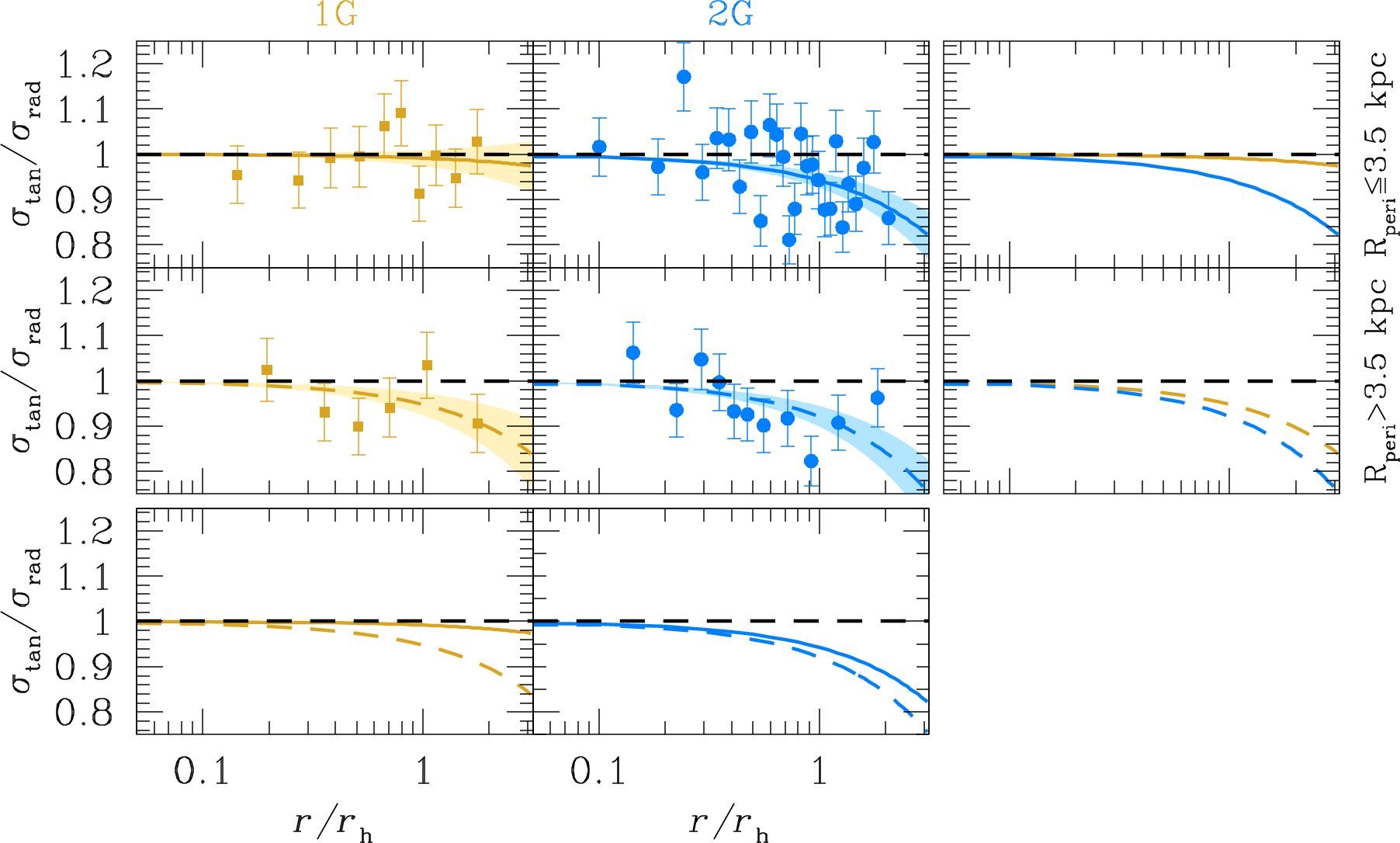}
\caption{Similarly to Fig.~\ref{fig:kin2}, we show the anisotropy as a function of distance from the centers of the clusters in unit of $r_{\rm h}$ for mPOPs in clusters with different ${\rm R_{peri}}$. The top and middle rows present the 1G (gold) and 2G (blue) anisotropy profiles for GCs with ${\rm R_{peri}} \le 3.5$ kpc and ${\rm R_{peri}} > 3.5$ kpc, respectively. The black, horizontal line is set to 1 (isotropic case). The colored lines are a straight-line fit to the data (see details in Fig~\ref{fig:kin1}). The comparison between the straight-line fit for mPOPs in GCs with the same ${\rm R_{peri}}$ is given in the rightmost panels, while between the same mPOP in GCs with different ${\rm R_{peri}}$ is highlighted in the bottom panels.\looseness=-4}
\label{fig:kin3}
\end{figure*}

\section{Dependence on the Galactic tidal field}\label{sec:rperi}

Previous studies \citep[e.g.,][]{2014MNRAS.443L..79V,2016MNRAS.455.3693T,2018MNRAS.475L..96B} have shown that the external tidal field of the host galaxy may play an important role in the early- and long-term evolution of the velocity anisotropy. In particular, the loss of stars in stronger tidal fields  preferentially affects stars on more radial orbits, causing a decrease of the radial anisotropy in the outer regions, and a gradual evolution towards a more isotropic (or even tangential) velocity distribution. We explore the possible role of the external tidal field as a function of perigalactic distances ($\rm R_{\rm peri}$) to quantify how much the Galactic tidal field affects the evolution of GCs.\looseness=-4

We have divided clusters into two groups with $\rm R_{\rm peri} > 3.5$ kpc and $\rm R_{\rm peri} \le 3.5$ kpc, respectively. The same value of the pericentric distance was adopted by \citet{2019MNRAS.487.3239Z} and \citet{2020MNRAS.491..515M} who explored the possible role of the Galactic tidal field on the fraction of 1G stars. Those studies found that the main correlation is between the fraction of 1G stars and the mass of the cluster and that, for a given value of the mass, clusters with larger pericentric distances tend to have larger 1G fractions.\looseness=-4

Two-body encounters progressively erase fingerprints of initial kinematic differences between mPOPs. Thus, we focused only on dynamically-young GCs (age/$t_{\rm h}$$<$7). This choice is also dictated by the properties of the GCs in our sample, given we have no dynamically-old and intermediate GCs with $\rm R_{\rm peri} > 3.5$ kpc. Red-RGB stars were excluded because there is only one type-II GC (NGC~6715) with age/$t_{\rm h}$$<$7 and large perigalactic distance.\looseness=-4

We show the anisotropy profiles for 1G and 2G stars for different $\rm R_{\rm peri}$ in Fig.~\ref{fig:kin3}. Our analysis suggests that the degree of kinematic anisotropy in 1G and 2G stars does depend on the perigalactic distance. The 1G stars are isotropic at all distances in our fields for $\rm R_{\rm peri} \le 3.5$ kpc, suggesting that the tidal field may have erased any initial anisotropy; the 2G population for clusters with the same pericentric distances, on the other hand, still displays some velocity anisotropy, in general agreement with what expected in models in which the 2G was initially more centrally concentrated than the 1G and thus less affected by the tidal field \citep{2021MNRAS.502.4290V}. In these clusters, we find that the average anisotropy for 1G and 2G groups at $r > 0.6\,r_{\rm h}$ are $1.01 \pm 0.03$ and $0.94 \pm 0.02$, respectively ($\sim$2$\sigma$ difference). In the group of clusters with $\rm R_{\rm peri} > 3.5$ kpc, both populations are still characterized by a radially-anisotropic velocity distribution, with an average anisotropy at $r > 0.6\,r_{\rm h}$ for 1G and 2G groups of $0.94 \pm 0.07$ and $0.91 \pm 0.03$, respectively. While the anisotropy of 2G stars is statistically significant at the 3$\sigma$ level, that of 1G sources is not as strong because the large error bars make the kinematics of this group consistent with the isotropic case at the $\sim$1$\sigma$ level. Similar results can be inferred by comparing the slopes of the corresponding least-squares straight-line fits. Our data show a marginal difference ($<$1$\sigma$) between 1G and 2G stars at large $r/r_{\rm h}$ for clusters in this group, but the possible larger anisotropy of the 2G population in the outer regions of clusters with large $\rm R_{\rm peri}$ needs to be investigated further over broader radial ranges.\looseness=-4

Finally, Table~\ref{tab:anisotropy} shows that the fraction of 1G stars in dynamically-young GCs with $\rm R_{\rm peri} < 3.5$ kpc is 0.26, while that in GCs with $\rm R_{\rm peri} > 3.5$ kpc is 0.33. This is qualitatively in agreement with the findings of \citet{2019MNRAS.487.3239Z} and \citet{2020MNRAS.491..515M}, i.e.,  GCs with larger $\rm R_{\rm peri}$ values tend to have larger 1G fractions.\looseness=-4

\section{Conclusions}

This paper presents the first homogeneous kinematic investigation of mPOPs in 56 GCs. We have focused on bright RGB stars, for which the mPOP tagging is clearer in chromosome maps, and measured the velocity dispersion of 1G and 2G stars. While 1G stars are, in general, kinematically isotropic at both inner and outer radii in our fields, 2G stars are isotropic at the center and progressively become more radially anisotropic further from the center of the cluster. This anisotropy is a reflection of the fact that the 2G stars have the same radial dispersions as the 1G stars, but much lower tangential dispersions. Our study confirms previous results obtained for specific GCs with \textit{Gaia} \citep{2018MNRAS.479.5005M,2020ApJ...889...18C,2020ApJ...898..147C} and \hst \citep{2010ApJ...710.1032A,2013ApJ...771L..15R,2015ApJ...810L..13B,2018ApJ...853...86B,2018ApJ...861...99L,2019ApJ...873..109L,2022ApJ...934..150L,2021MNRAS.506..813D}. Our findings are also in general agreement with the theoretical predictions of models that follow the dynamical evolution of mPOPs and show that these properties are expected in systems in which 2G stars formed more centrally concentrated than 1G stars.\looseness=-4

Using our sample, we also find possible indications that the Galactic tidal fields affect the kinematic properties of 1G and 2G stars. Specifically, we show that the anisotropy of 1G and 2G stars depends on the perigalactic distance $\rm R_{\rm peri}$ of the host cluster. Systems with large $\rm R_{\rm peri}$ experience, on average, a weaker tidal field and their stars are able to preserve a (stronger) radial anisotropy than GCs with pericentric distances in the innermost regions of the Galaxy \citep[see also][for the possible effect of $\rm R_{\rm peri}$ on the fraction of 1G stars]{2019MNRAS.487.3239Z,2020MNRAS.491..515M}.\looseness=-4

Although these results are not conclusive, due to the limited sample and limited radial coverage, our analysis is a step forward towards a complete understanding of the mPOP phenomenon. This initial study provides further motivation for new and deeper surveys with \hst, \textit{JWST} and, in the future, the \textit{Nancy Grace Roman Space Telescope}, which will be essential to extend the investigation in the almost-uncharted outskirts of GCs \citep{2019JATIS...5d4005W,2019astro2020T.173B}.\looseness=-4

\section*{Acknowledgments}

The authors thank the anonymous referee for the detailed suggestions that improved the quality of our work. ML and AB acknowledges support from GO-13297, GO-15857 and GO-16298. EV acknowledges support from NSF grant AST-2009193. APM acknowledgesfunding from the European Research Council (ERC) under the European Union's Horizon 2020 research innovation programme (Grant Agreement ERC-StG 2016, No 716082 'GALFOR', PI: Milone, \href{http://progetti.dfa.unipd.it/GALFOR}{http://progetti.dfa.unipd.it/GALFOR}). AA acknowledges support from the Ministerio de Ciencia e Innovaci\'on of Spain (grant PID2020-115981GB-I00), from the Instituto de Astrof\'isica de Canarias (grant P/309403) and from Khalifa University of Abu Dhabi through project CIRA-2021-65 8474000413. BB acknowledges partial financial support from FAPESP, CNPq and CAPES - financial code 001. AS is grateful to the Bjorn Lamborn Endowment for support while some of this work was being done. ML thanks LS, GL and LL for their support to complete this project.\looseness=-4

This research was pursued in collaboration with the HSTPROMO (High-resolution Space Telescope PROper MOtion) collaboration, a set of projects aimed at improving our dynamical understanding of stars, clusters and galaxies in the nearby Universe through measurement and interpretation of proper motions from \textit{HST}, \textit{Gaia}, and other space observatories. We thank the collaboration members for the sharing of their ideas and software.\looseness=-4

Based on observations with the NASA/ESA \textit{HST}, obtained at the Space Telescope Science Institute, which is operated by AURA, Inc., under NASA contract NAS 5-26555. This research made use of \texttt{astropy}, a community-developed core \texttt{python} package for Astronomy \citep{astropy:2013, astropy:2018, 2022ApJ...935..167A}, and \texttt{emcee} \citep{2013PASP..125..306F}.\looseness=-4

\bibliography{mPOPs_RGB_kin}{}
\bibliographystyle{aasjournal}

\end{document}